%% file: main.tex
\pdfoutput=1
\documentclass[11pt,letterpaper]{article}

\newif\ifshowsummary
\showsummaryfalse

\usepackage[T1]{fontenc}
\usepackage[utf8]{inputenc}
\usepackage{XCharter}
\usepackage[xcharter,vvarbb]{newtxmath}
\usepackage[scale=0.92]{sourcesanspro}
\usepackage[varqu,scaled=0.9]{zi4}
\usepackage[final,protrusion=true,expansion=true,tracking=true]{microtype}
\SetTracking{encoding=*, shape=sc}{50}

\usepackage[letterpaper, textwidth=6.25in, textheight=9in,
            top=1in, headsep=0.3in, footskip=0.45in]{geometry}
\usepackage{amsmath}
\usepackage{xcolor}
\usepackage{graphicx}
\usepackage{booktabs}
\usepackage{tabularx}
\usepackage{array}
\usepackage[shortlabels]{enumitem}
\usepackage{tikz}
\usetikzlibrary{arrows.meta, positioning, shapes.geometric, calc, fit, backgrounds}
\usepackage[most]{tcolorbox}
\usepackage{algorithm}
\usepackage[noend]{algpseudocode}
\usepackage[font=small, labelfont=bf, labelsep=period,
            justification=justified, singlelinecheck=false,
            skip=6pt]{caption}
\usepackage{titlesec}
\usepackage{fancyhdr}
\usepackage[round]{natbib}
\usepackage{xurl}

\definecolor{kdink}{HTML}{1A1A1A}
\definecolor{kdgrey}{HTML}{6E6E6E}
\definecolor{kdgreen}{HTML}{2E5E55}
\definecolor{kdgreenbg}{HTML}{EEF2F1}
\definecolor{kdnavy}{HTML}{3A3A3A}
\definecolor{kdnavybg}{HTML}{F4F4F4}
\definecolor{kdamber}{HTML}{A85A24}
\definecolor{kdamberbg}{HTML}{FAF0E8}
\definecolor{kdgold}{HTML}{7C7C7C}
\definecolor{kdgoldbg}{HTML}{F8F8F8}
\definecolor{kdred}{HTML}{5A5A5A}
\definecolor{kdredbg}{HTML}{F3F3F3}
\definecolor{kdlink}{HTML}{1D3A6E}

\titleformat{\section}{\normalfont\large\bfseries}{\thesection}{0.75em}{}
\titleformat{\subsection}{\normalfont\normalsize\bfseries}{\thesubsection}{0.75em}{}
\titleformat{\subsubsection}{\normalfont\normalsize\itshape}{\thesubsubsection}{0.75em}{}
\titleformat{\paragraph}[runin]{\normalfont\normalsize\itshape}{}{0pt}{}[.]
\titlespacing*{\section}{0pt}{2.4ex plus .6ex minus .3ex}{1.1ex plus .2ex}
\titlespacing*{\subsection}{0pt}{2.0ex plus .5ex minus .3ex}{0.8ex plus .2ex}
\titlespacing*{\subsubsection}{0pt}{1.6ex plus .4ex minus .2ex}{0.6ex plus .2ex}
\titlespacing*{\paragraph}{0pt}{1.0ex plus .3ex minus .2ex}{0.5em}

\setlist{itemsep=1pt, topsep=4pt, parsep=0pt, partopsep=0pt}
\setlist[itemize]{leftmargin=1.5em}
\setlist[enumerate]{leftmargin=2em}

\fancypagestyle{plain}{\fancyhf{}\fancyfoot[C]{\footnotesize\thepage}}

\algrenewcommand\algorithmicrequire{\textit{Input:}}
\algrenewcommand\algorithmicensure{\textit{Output:}}
\algrenewcommand\alglinenumber[1]{\footnotesize #1}
\algnewcommand{\LineComment}[1]{\State \textcolor{kdgrey}{\(\triangleright\) \textit{#1}}}

\usepackage[pdfusetitle]{hyperref}
\hypersetup{
  colorlinks=true, linkcolor=kdlink, citecolor=kdlink, urlcolor=kdlink,
  pdftitle={KumbhDoot: A Scale-Ready, LLM-Bounded Architecture for Mass-Gathering Public-Service Assistants},
  pdfauthor={Saurabh Sakalkar, Abhishek Singh, Ramesh Raskar},
  pdfsubject={Information retrieval; public-service assistants; retrieval-augmented generation},
  pdfkeywords={semantic cache, retrieval-augmented generation, intent routing, offline-first, mass gatherings, Kumbh Mela}
}
\usepackage[capitalise,noabbrev,nameinlink]{cleveref}
\crefname{requirement}{Requirement}{Requirements}

\newcommand{\KD}{KumbhDoot}
\newcommand{\band}[1]{Band~#1}
\newcommand{\R}[1]{\textsf{R#1}}
\newcommand{\vect}[1]{\mathbf{#1}}
\DeclareMathOperator*{\argmax}{arg\,max}
\DeclareMathOperator{\topk}{top\text{-}k}

\begin{document}

\thispagestyle{plain}
\begin{center}
  {\Large\bfseries
   \KD: A Scale-Ready, LLM-Bounded Architecture\\[2pt]
   for Mass-Gathering Public-Service Assistants\par}
  \vspace{1.3em}
  {\normalsize
   Saurabh Sakalkar\textsuperscript{1}\qquad
   Abhishek Singh\textsuperscript{1}\qquad
   Ramesh Raskar\textsuperscript{2}\par}
  \vspace{0.5em}
  {\small
   \textsuperscript{1}Project Nanda\qquad
   \textsuperscript{2}Massachusetts Institute of Technology\par}
\end{center}
\vspace{0.8em}

\begin{center}
\begin{minipage}{0.86\textwidth}
\small
\setlength{\parindent}{1.5em}
\centerline{\textsc{abstract}}
\vspace{0.5ex}
\noindent Mass religious gatherings such as the Kumbh Mela bring tens of millions of people to a
single region within a few weeks and create heavy, repetitive, multilingual demand for information,
part of which concerns personal safety. Conversational assistants are usually built by sending every
query to a large language model (LLM). In this setting that design is expensive at scale, slow on
emergency paths, liable to hallucinate facts on which physical safety depends, and unavailable when
the network fails. We describe \KD, a pilgrim assistant for the Nashik Simhastha Kumbh Mela, in which
semantic similarity over approved content is the default mechanism and an LLM is called only when
retrieval cannot produce a correct answer. A single embedding-indexed store, which we call the
semantic cache, serves as the system's retrieval primitive for intent routing, answer caching,
offline lookup, and agent retrieval. A hand-written three-tier orchestrator operates on this store,
so that each call to a model is the outcome of an explicit gate rather than a framework default. We
present the architecture, an analysis of its per-query cost in terms of how traffic divides between
retrieval-only and generative paths, and a statement of which query types similarity can and cannot
serve. We argue that, for bounded, safety-critical domains with unreliable connectivity, a
similarity-first design with gated LLM use is a better fit than the conventional LLM-centric pattern.
We do not yet report a controlled evaluation; \cref{sec:limitations} sets out the measurements that
the claims require.

\vspace{0.8ex}
\end{minipage}
\end{center}

\section{Introduction}
\label{sec:intro}

An information service for a Kumbh-scale event must work under conditions that seldom coincide.
Millions of people use it at once, mostly on inexpensive phones and over congested mobile networks.
They ask questions in several languages, often mixing them within a sentence and often by voice. Most
questions are routine, such as bathing dates, the nearest ghat, or a helpline number, but some report
a lost child, a medical problem, or a dangerous crowd. For the latter, a slow or wrong answer is a
safety failure.

The usual way to build a conversational assistant today is to pass each utterance to an LLM for both
understanding and generation. We call this the \emph{LLM-default} pattern. It is convenient for
prototyping, but at event scale it has four problems that compound one another. Repeated questions are
regenerated, and are paid for, every time they are asked. Generation adds latency to paths that must be
immediate. Free-form generation over facts such as ghat access, medical contacts, and ritual schedules
can produce plausible but false statements in precisely the cases where they cause harm. And a
cloud-hosted model is unavailable exactly when crowds saturate the network.

Our starting point is that most of the work in such a service is retrieval rather than generation.
Representing text as vectors and ranking candidates by cosine similarity is an old problem in
information retrieval; the operation is inexpensive, deterministic, and easy to audit. \KD{} uses it
as the default path for the large majority of queries and reserves the LLM for requests that require
composition, that fall below a retrieval-confidence threshold, or that depend on live external data.

The techniques we use (dense sentence embeddings, hybrid lexical and dense retrieval, cross-encoder
reranking, and nearest-neighbor classification over embeddings) are established. Our contribution lies
in how they are combined and constrained for a public-service setting:
\begin{enumerate}[(i)]
  \item a design in which one retrieval primitive, cosine similarity over partitions of a single
        embedding-indexed store, provides the knowledge base, intent classifier, response cache, offline
        bundle, and per-agent memory (\cref{sec:semantic-cache});
  \item an escalation policy that orders methods by cost, calls an LLM only when an explicit gate
        fires, and validates the citations in any generated answer against the retrieved records
        (\cref{sec:ladder}), together with a hand-written three-tier orchestrator built on the same
        primitive (\cref{sec:orchestration});
  \item an analysis that expresses expected per-query cost in terms of the share of traffic that must
        escalate to generation (\cref{sec:cost}); and
  \item an explicit account of the query classes that similarity can and cannot serve, with an
        evaluation plan for the claims that remain unmeasured (\cref{sec:bands,sec:limitations}).
\end{enumerate}

\Cref{sec:requirements} states the deployment setting and the requirements it imposes.
\Cref{sec:related} reviews related work. \Cref{sec:design} describes the system, \cref{sec:cost}
analyzes its cost, and \cref{sec:bands} discusses the limits of similarity. \Cref{sec:limitations}
lists limitations and planned evaluation, and \cref{sec:conclusion} concludes.

\section{Setting and Requirements}
\label{sec:requirements}

\KD{} is designed for the Nashik Simhastha Kumbh Mela. Its users are pilgrims, who arrive over several
weeks, speak mainly Marathi, Hindi, or English, and rely on phones with limited compute and
intermittent connectivity. Its content is supplied by the event administration: official advisories,
facility locations, maps, ritual schedules, and emergency contacts. From this setting we derive six
requirements, numbered \R{1}--\R{6} for reference in later sections; \cref{tab:requirements} pairs each
with the mechanism that addresses it.

The volume of traffic is the most immediate constraint. Because the same questions recur across
millions of users, an assistant that regenerates every answer turns a largely cacheable workload into a
recurring per-query expense. Cost should instead grow with the number of questions that are new in
substance, as opposed to new in wording (\R{1}). Within that traffic, a small fraction of messages
report a lost child or ask for an ambulance. Such messages must receive deterministic guidance at
once, without waiting on multi-step generation (\R{2}).

A second group of requirements concerns what the system may say and under what conditions it can say
it. Statements about routes, medical facilities, and ritual schedules must be drawn from approved
records; in this domain grounding is a safety control as much as a quality
control~\citep{shuster2021retrieval}, and we therefore require it of every answer (\R{3}). Dense crowds
also degrade mobile networks, often at the moments when information is most needed, so the core of
the service has to function on the device without a connection (\R{4}). A design whose every answer
depends on a cloud-hosted model cannot satisfy this requirement.

The last two requirements concern data. A pilgrim's location, health status, family composition, and
stage of journey are sensitive, and routing every utterance to a remote model would expose precisely
the information that ought to remain on the device or be coarsened before it leaves (\R{5}). The event
administration, for its part, needs aggregate signals, such as where crowds are concentrating and which
concerns are emerging, and has no need for individual conversations or per-user tracking (\R{6}).

\begin{table}[t]
  \centering
  \caption{Requirements of the deployment setting and the mechanisms that address them.}
  \label{tab:requirements}
  \small
  \begin{tabularx}{\textwidth}{@{}l>{\raggedright\arraybackslash}X>{\raggedright\arraybackslash}Xl@{}}
    \toprule
    & Requirement & Principal mechanism & Section \\
    \midrule
    \R{1} & Cost proportional to novelty & Paraphrase-aware response cache; answers formatted from records without an LLM & \ref{sec:semantic-cache}, \ref{sec:ladder} \\
    \R{2} & Immediate emergency handling & Keyword-triggered, rule-based fast path ahead of intent routing & \ref{sec:intent} \\
    \R{3} & Grounding in approved content & InfoBins with source and validity metadata; coded formatters; citation validation & \ref{sec:infobins}, \ref{sec:ladder} \\
    \R{4} & Offline operation & On-device SQLite bundle with FTS5 and pre-computed embeddings & \ref{sec:offline} \\
    \R{5} & Data minimization & Local-first storage, coarse and consented location, bucketed attributes & \ref{sec:limitations} \\
    \R{6} & Aggregate operational insight & Observability across LLM and non-LLM paths; aggregate dashboards & \ref{sec:runtime} \\
    \bottomrule
  \end{tabularx}
\end{table}

\section{Related Work}
\label{sec:related}

\paragraph{Lexical and dense retrieval}
Lexical scoring rests on the vector space model~\citep{salton1975vector} and on term
weighting~\citep{sparckjones1972statistical}; latent semantic analysis~\citep{deerwester1990indexing} introduced
low-dimensional dense representations; \citet{manning2008introduction} give a standard treatment. Word
embeddings~\citep{mikolov2013efficient} and, later, sentence encoders trained with siamese or
contrastive objectives~\citep{reimers2019sentence,gao2021simcse} made it practical to compare whole
queries by cosine similarity, which is the granularity at which \KD{} operates. Cross-encoders improve
precision by rescoring a candidate set produced by a cheaper first stage~\citep{nogueira2019passage,cheng2026enhancing},
and reciprocal rank fusion offers a simple, robust way to combine lexical and dense
rankings~\citep{cormack2009reciprocal}. Throughout this line of work the representations have improved
while the comparison operation has remained essentially the same.

\paragraph{Retrieval-augmented generation}
Retrieval-augmented generation conditions a language model on retrieved passages to improve factuality
on knowledge-intensive tasks~\citep{lewis2020retrieval,guu2020retrieval}, typically with a dense
retriever~\citep{karpukhin2020dense}, and retrieval has been shown to reduce hallucination in
dialogue~\citep{shuster2021retrieval}. In these systems a model generates every answer and retrieval
supplies its context. \KD{} reverses the default: most answers are assembled from retrieved records
without a model, and generation is the exception.

\paragraph{LLM agents}
Agent frameworks interleave model reasoning with tool calls~\citep{yao2022react}. They are flexible,
but control flow, including intent classification and task decomposition, is usually delegated to the
model, so that each step incurs a call and its decision is difficult to audit. \KD{} keeps
orchestration in code and reaches a model only through explicit gates.

\paragraph{Caching, offline operation, and on-device models}
Practitioner systems cache prompts and responses to avoid repeated model calls, keep application data
in offline-first mobile stores that synchronize on reconnection, and trace LLM calls for latency and
cost. \KD{} applies each of these ideas more broadly: its cache matches paraphrases rather than exact
strings, its offline tier performs intent routing and retrieval as well as storage, and its
observability covers non-LLM paths. Work on distillation~\citep{hinton2015distilling},
compression~\citep{han2015deep}, and on-device learning~\citep{mcmahan2016communication} has made small
models practical on handsets; \KD{} uses such components where they are available but does not assume
them. Large language models~\citep{brown2020language} provide the generative capability that \KD{}
uses for the residual class of queries described in \cref{sec:bands}.

\medskip
\noindent What distinguishes \KD{} from these lines of work is scope rather than any single technique.
It is an event-specific assistant with bounded knowledge, strict fallbacks, and an offline tier, and
that scope is what makes a similarity-first design viable.

\section{System Design}
\label{sec:design}

\subsection{Overview}
\label{sec:runtime}

The runtime is a stack of layers, each of which answers what it can before passing the query upward
(\cref{fig:runtime}). From the bottom, a curated content layer holds official advisories, facility
details, maps, schedules, and emergency contacts. An InfoBin layer turns this content into versioned,
searchable records (\cref{sec:infobins}), and a compact partition of those records is shipped to the
device as a local semantic cache (\cref{sec:offline}). An intent router combines rules, light
normalization, and nearest-neighbor matching with confidence scores (\cref{sec:intent}). Bounded domain
agents serve food, travel, safety, accommodation, health, emergency, ritual, profile, and crowd
queries (\cref{sec:orchestration}). A response composer formats answers from approved records and calls
a small model or an LLM only when the escalation policy requires it (\cref{sec:ladder}). An
observability layer spans the stack and records latency, cost, cache hits, model use, sources, safety
flags, and fallback paths for every request, including those that never reach a model.

\begin{figure}[t]
  \centering
  \input{figures/fig-runtime.tex}
  \caption{Layered runtime. A query is answered by the lowest layer able to answer it reliably and moves
    upward only when that layer cannot. Observability spans all layers and covers both LLM and non-LLM
    paths.}
  \label{fig:runtime}
\end{figure}

\subsection{InfoBins}
\label{sec:infobins}

Informational apps usually present curated content as static cards. An agent needs the same content
in a form that can be matched against natural-language queries, checked for currency, and withdrawn
when it becomes wrong. We call such a record an \emph{InfoBin} and write it as
\begin{equation}
  b = (c,\ \mu,\ \vect{v},\ \sigma),
\end{equation}
where $c$ is the approved content; $\mu$ is metadata comprising domain, language, source authority,
zone, validity interval, and time of last update; $\vect{v} \in \mathbb{R}^{d}$ is a unit-normalized
embedding of the content; and $\sigma$ holds safety attributes, namely the source label, freshness
status, revocation status, and escalation rule. The first question the runtime asks of any query is
whether an approved InfoBin already answers it; escalation follows only when none does.

\subsection{The semantic cache}
\label{sec:semantic-cache}

Let $\phi$ denote a sentence encoder whose outputs are unit-normalized, so that cosine similarity
reduces to an inner product. The semantic cache is a collection of partitions $\mathcal{S} = \{P_1,
\dots, P_K\}$, where each partition $P_k = \{(\vect{v}_j, \pi_j)\}$ pairs key embeddings with payloads.
All retrieval in the system is an instance of one operation,
\begin{equation}
  \label{eq:topk}
  \operatorname{Retrieve}(q, P_k) = \topk_{(\vect{v}, \pi) \in P_k} \ \langle \phi(q), \vect{v} \rangle,
\end{equation}
optionally fused with a lexical ranking and reranked (\cref{sec:ladder}). Five functions that are
usually built as separate components differ only in the partition queried (\cref{tab:partitions,fig:cache}).

\begin{table}[t]
  \centering
  \caption{Five functions expressed as retrieval against different partitions of the semantic cache.}
  \label{tab:partitions}
  \small
  \begin{tabularx}{\textwidth}{@{}l>{\raggedright\arraybackslash}X>{\raggedright\arraybackslash}X@{}}
    \toprule
    Function & Keys & Payload \\
    \midrule
    Knowledge base     & Embeddings of InfoBin content            & InfoBin records \\
    Intent classifier  & Embeddings of labeled exemplar queries   & Intent and domain labels \\
    Response cache     & Embeddings of previously answered queries & Approved answers \\
    Offline bundle     & Device-resident subset of InfoBin embeddings, with an FTS5 index & InfoBin records \\
    Per-agent memory   & Embeddings of records in an agent's domain partition & Records scoped to that agent \\
    \bottomrule
  \end{tabularx}
\end{table}

\begin{figure}[t]
  \centering
  \input{figures/fig-cache.tex}
  \caption{One retrieval primitive in place of five components. Each function is cosine similarity
    against a different partition of one embedding-indexed store; the variety lies in the partition,
    not the algorithm.}
  \label{fig:cache}
\end{figure}

The semantic cache is a logical layer rather than a single database, and it is realized differently
on the server and on the device (\cref{tab:tiers}). On the server, domain partitions are held in memory
and in cache-backed stores, including a Redis store for paraphrase-aware query-to-answer caching, with
higher-dimensional embeddings that are updated as content changes. On the device, a curated partition
is shipped as a SQLite bundle of approximately 700\,KB that combines a full-text (FTS5) index with
lower-dimensional embeddings stored as binary blobs, so that hybrid retrieval requires no network call.
Both tiers expose the abstraction of \cref{eq:topk}. The tiers do not share a physical store, and live
synchronization from the server indexes to the device bundle is not yet implemented: the bundle is
built at release time.

\begin{table}[t]
  \centering
  \caption{Realization of the semantic cache on the two tiers.}
  \label{tab:tiers}
  \small
  \begin{tabularx}{\textwidth}{@{}l>{\raggedright\arraybackslash}X>{\raggedright\arraybackslash}X@{}}
    \toprule
    & Server & Device \\
    \midrule
    Storage           & In-memory partitions; Redis for query-to-answer pairs & SQLite bundle ($\approx$700\,KB) \\
    Embeddings        & Higher-dimensional; updated with live content & Lower-dimensional; pre-computed, stored as blobs \\
    Lexical index     & ---                                  & SQLite FTS5 \\
    Retrieval         & Dense retrieval, fusion, reranking   & FTS5 and dense scores combined by rank fusion \\
    Query embedding   & Server-side encoder                  & Quantized on-device encoder if present; otherwise word-overlap matching against pre-computed query embeddings \\
    Content updates   & Continuous                           & At build time (live synchronization not yet implemented) \\
    \bottomrule
  \end{tabularx}
\end{table}

\subsection{Intent routing}
\label{sec:intent}

Many agent frameworks classify intent by prompting a model~\citep[e.g.,][]{yao2022react}, which adds a
call, latency, and an opaque decision to every request. \KD{} treats intent classification as
nearest-neighbor search. Let $\mathcal{E} = \{(\vect{e}_i, y_i)\}_{i=1}^{n}$ be a set of $n \approx 520$
exemplar queries labeled with one of 26 intents, written in English, Hindi, and Marathi and embedded
once at start-up. For a query $q$, the router computes
\begin{equation}
  \hat{\imath} = \argmax_{i} \ \langle \phi(q), \vect{e}_i \rangle, \qquad
  \hat{y} = y_{\hat{\imath}}, \qquad
  s = \langle \phi(q), \vect{e}_{\hat{\imath}} \rangle,
\end{equation}
which for all exemplars is a single matrix--vector product. The predicted intent $\hat{y}$ determines
the domain, and the score $s$ is retained as an auditable confidence. The step uses no generative
model, although the encoder still runs.

Two components surround this step. Before it, the router normalizes transliteration and common
speech-recognition errors and checks the query against a list of emergency phrases; a match routes the
query to a rule-based fast path. After it, a separate heuristic estimates query complexity from the
number of domain keywords, the presence of conjunctions in any of the three languages, and query
length. Complexity, not intent, decides whether the orchestrator of \cref{sec:orchestration} is
engaged.

\subsection{Escalation policy}
\label{sec:ladder}

\Cref{alg:ladder} and \cref{fig:ladder} give the order in which the runtime tries its methods. A query
is first matched against the response cache; a confident hit is returned through a coded formatter. On
a miss, the router assigns a domain, and hybrid retrieval over that domain's partition produces a
candidate set. When retrieval is confident and the request is single-domain, the records are formatted
into a cited answer without a model. An LLM is called only when an escalation gate fires: the request
needs multi-part synthesis, retrieval confidence falls below threshold, or the answer depends on live
or external data. Generation is then restricted to the retrieved records, and a validation step removes
any citation that does not correspond to a record the retriever returned, so that identifiers invented
by the model do not reach the user.

\begin{algorithm}[t]
  \caption{Escalation policy for a single query. Thresholds $\tau_{\mathrm{c}}$ and $\tau_{\mathrm{r}}$
    govern the cache and retrieval gates.}
  \label{alg:ladder}
  \small
  \begin{algorithmic}[1]
    \Require query $q$; semantic cache $\mathcal{S}$; exemplars $\mathcal{E}$
    \Ensure an answer with its sources and freshness status
    \State $\vect{z} \gets \phi(q)$
    \State $(a, s_{\mathrm{c}}) \gets \operatorname{Retrieve}_1(\vect{z}, P_{\mathrm{cache}})$
    \If{$s_{\mathrm{c}} \geq \tau_{\mathrm{c}}$} \Return $\operatorname{Format}(a)$
      \Comment{\textcolor{kdgrey}{no LLM}}
    \EndIf
    \State $q' \gets \operatorname{Normalize}(q)$
      \Comment{\textcolor{kdgrey}{transliteration, ASR errors}}
    \If{$\operatorname{EmergencyMatch}(q')$} \Return $\operatorname{EmergencyResponse}(q')$
    \EndIf
    \State $(\hat{y}, s) \gets \operatorname{RouteIntent}(q', \mathcal{E})$
    \State $R \gets \operatorname{HybridRetrieve}(q', P_{\mathrm{dom}(\hat{y})})$
      \Comment{\textcolor{kdgrey}{lexical + dense, fused, reranked}}
    \If{$\neg\operatorname{Complex}(q') \wedge \operatorname{conf}(R) \geq \tau_{\mathrm{r}} \wedge \neg\operatorname{NeedsLive}(q')$}
      \State \Return $\operatorname{FormatFromRecords}(R)$
        \Comment{\textcolor{kdgrey}{no LLM}}
    \EndIf
    \State $A \gets \operatorname{BoundedLLM}(q', R)$
      \Comment{\textcolor{kdgrey}{generation restricted to $R$}}
    \State \Return $\operatorname{ValidateCitations}(A, R)$
      \Comment{\textcolor{kdgrey}{drop citations not in $R$}}
  \end{algorithmic}
\end{algorithm}

\begin{figure}[t]
  \centering
  \input{figures/fig-ladder.tex}
  \caption{Escalation ladder. Methods are tried in order of cost. Paths drawn in the dark accent return
    without a generative model; the warm-colored path is the only one that calls an LLM.}
  \label{fig:ladder}
\end{figure}

\subsection{Orchestration}
\label{sec:orchestration}

Requests that span domains are handled by three tiers of agents that share the retrieval primitive
(\cref{fig:tiers}). An \emph{orchestrator} decomposes the request into sub-needs with dependency edges
and schedules them in waves: independent sub-needs run concurrently, and dependent ones wait for, and
receive, the results of their prerequisites. \emph{Domain agents}, one each for health, travel,
commerce, crowd, identity, governance, curated Kumbh content, and other domains, retrieve from their own
partitions. \emph{Record-level sub-agents} turn retrieved items into structured answers, annotating
each with whether it is actionable, its card type, whether it concerns an emergency, and whether it
matches a name in the query.

\begin{figure}[t]
  \centering
  \input{figures/fig-tiers.tex}
  \caption{Three-tier orchestration. All tiers retrieve through the same primitive, each over its own
    partition; only four of the domain agents are shown. Sub-agents annotate retrieved items with the
    fields \texttt{\_actionable}, \texttt{\_card\_type}, \texttt{\_is\_emergency}, and
    \texttt{\_name\_match}.}
  \label{fig:tiers}
\end{figure}

The orchestration logic is written by hand for this domain rather than delegated to a general-purpose
agent framework. Because every agent consults the semantic cache before anything else, a model is
reached only under the conditions of \cref{alg:ladder}, and the path by which each answer was produced
can be read from the trace. A further consequence is that adding a domain is mostly a matter of data:
curated records are loaded into a new partition and the domain is registered. New orchestration code
is needed only when the shared schema or routing rules do not already accommodate the domain.

\subsection{Offline operation}
\label{sec:offline}

The device tier is designed to remain useful without a connection. Emergency contacts, key facilities,
the user's saved itinerary, basic maps, frequently asked questions, phrasebooks, and previously
synchronized InfoBins are stored locally. When the network is unavailable, voice input falls back to
text, agentic answers fall back to approved static guidance, and live data is shown with the time it
was last updated. Non-critical requests are queued until the connection returns, and emergency updates
are synchronized first. Every answer is labeled as \emph{available offline}, \emph{last updated}, or
\emph{pending refresh}, so that users can judge how current it is. Retrieval on the device fuses FTS5
results with dense scores by rank fusion, as summarized in \cref{tab:tiers}.

\section{Cost Analysis}
\label{sec:cost}

In a pipeline where every stage is an LLM call, cost accrues at intent classification, complexity
assessment, decomposition, per-domain reasoning, and synthesis. A conventional retrieval-augmented
pipeline removes the model from retrieval but still generates every answer. In \KD{} each stage has a
deterministic implementation, and the model is a fallback for synthesis (\cref{tab:stages}).

\begin{table}[t]
  \centering
  \caption{Where a generative model is invoked in three pipeline designs.}
  \label{tab:stages}
  \small
  \begin{tabularx}{\textwidth}{@{}l>{\raggedright\arraybackslash}X>{\raggedright\arraybackslash}X>{\raggedright\arraybackslash}X@{}}
    \toprule
    Stage & LLM-default & Conventional RAG & \KD \\
    \midrule
    Intent classification & LLM & --- & Nearest-neighbor over exemplars \\
    Complexity assessment & LLM & --- & Keyword and length heuristic \\
    Decomposition         & LLM & --- & Hand-written orchestrator \\
    Retrieval             & LLM-directed tools & Dense retriever & Hybrid retrieval with reranking \\
    Answer synthesis      & LLM & LLM & Coded formatter; LLM only when a gate fires \\
    \bottomrule
  \end{tabularx}
\end{table}

The deterministic stages are inexpensive. Routing against a fixed exemplar set is one matrix--vector
product; a response-cache lookup is a nearest-neighbor query; hybrid retrieval combines vector
arithmetic, a lexical index, and a small local reranker. None of these consumes generative tokens. The
query embedding does consume compute, so the accurate statement is that common paths use no generative
tokens, not that they are free. On the device, where content vectors are pre-computed, even the
content-side embedding is avoided.

Let $p_A$, $p_B$, and $p_C$ be the fractions of traffic answered in Bands~A, B, and~C
(\cref{sec:bands}), with $p_A + p_B + p_C = 1$, and let $c_A$, $c_B$, and $c_C$ be the corresponding
per-query costs excluding the query embedding, whose cost $c_{\mathrm{emb}}$ is paid on every
server-side path. The expected cost of a query is
\begin{equation}
  \label{eq:cost}
  \mathbb{E}[C] = c_{\mathrm{emb}} + p_A\,c_A + p_B\,c_B + p_C\,c_C,
  \qquad c_A \le c_B \ll c_C,
\end{equation}
where the ordering holds because Bands~A and~B use no generative model. In an LLM-default design every
query costs on the order of $c_C$. The advantage of the similarity-first design therefore depends
chiefly on $p_C$, the share of traffic that must escalate.

That share should fall as the response cache fills. The first occurrence of a question costs one
embedding and one retrieval; later paraphrases, in any supported language, are answered from the cache
whenever they fall within its similarity threshold. If pilgrimage queries are concentrated on a
relatively small set of recurring needs, as we expect, the fraction answerable without a model grows
with use, and the marginal cost of a common query approaches that of a similarity computation.
Generation is then paid for only on the tail of novel, multi-step, or externally sourced requests. This
is an argument about the structure of the cost, not a measurement; \cref{sec:limitations} describes how
$p_C$ and the band costs will be estimated.

\section{What Similarity Can and Cannot Do}
\label{sec:bands}

In a bounded, curated domain, similarity search with coded formatting suffices for a large class of
needs: intent classification over a fixed label set, retrieval that tolerates paraphrase,
FAQ-style answers, and structured factual responses drawn from approved records. For these needs a
generative model adds latency, cost, and risk of error without adding correctness.

Several needs lie outside this class. A query about something absent from the corpus produces
low-confidence retrieval and must be deflected or escalated; retrieval cannot return information that
the corpus does not contain. The closest record is not necessarily the intended one, particularly near
intent boundaries. Planning under constraints that span domains requires orchestration and, often,
generated prose. Coded formatters produce accurate structured output but not fluent narrative.
Transactions such as booking or identity verification require tool-using agents. Multilingual
encoders reduce, but do not remove, degradation from code-mixing, dialect, and speech-recognition
errors.

These observations divide the workload into three bands (\cref{fig:bands}). In \band{A}, a query is
answered from the response cache or by direct retrieval, as when a user asks where a ghat is or when
the bathing dates fall. In \band{B}, coded orchestration combines several records into a structured
answer, as when a user asks for hospitals with emergency care near a locality. In \band{C}, the system
escalates to LLM-backed synthesis, as when an elderly traveler asks for a personalized plan for the day
of arrival. The aim is to answer each query in the cheapest band that can answer it correctly and to
make every escalation to \band{C} the result of an explicit decision.

\begin{figure}[t]
  \centering
  \input{figures/fig-bands.tex}
  \caption{Three workload bands. Each query should be answered in the cheapest band that can answer it
    correctly. Hallucination is reduced rather than eliminated, because \band{C} still calls a model.}
  \label{fig:bands}
\end{figure}

The same reasoning bounds what can be claimed about hallucination. Because most answers are assembled
by coded formatters from curated records, the opportunities for unconstrained generation are fewer
than in a design that calls a model at every step. They are not zero: \band{C} still calls a model, and
retrieval can surface a record that is itself wrong. Citation validation removes one class of
fabrication but does not guarantee correctness.

\section{Limitations and Planned Evaluation}
\label{sec:limitations}

This paper describes an architecture and argues for it analytically; it does not report a controlled
evaluation. \Cref{tab:evaluation} summarizes the measurements the claims require, and the paragraphs
below discuss the limitations of the current design.

\begin{table}[t]
  \centering
  \caption{Claims that remain to be measured and the planned protocol for each.}
  \label{tab:evaluation}
  \small
  \begin{tabularx}{\textwidth}{@{}>{\raggedright\arraybackslash}p{0.27\textwidth}>{\raggedright\arraybackslash}X@{}}
    \toprule
    Claim & Planned measurement \\
    \midrule
    Most traffic needs no generative model & Share of queries answered in Bands A and B on a representative query distribution, with full per-step tracing \\
    Cost and latency by band & Tokens, cost, and latency per step on a tiered benchmark of simple, multi-record, and multi-domain queries, each paired with paraphrases \\
    Advantage over alternatives & Comparison against both an LLM-at-every-step pipeline and a conventional RAG pipeline that uses a model only for synthesis \\
    Multilingual routing accuracy & Accuracy and threshold calibration on a labeled test set in English, Hindi, Marathi, and code-mixed input, including speech-recognition noise \\
    Offline usefulness & Minimum useful bundle size and on-device hit rate, measured by replaying representative traffic \\
    \bottomrule
  \end{tabularx}
\end{table}

The choice of baselines matters for the comparison in the third row of \cref{tab:evaluation}. A
pipeline that forces a model through every stage represents a worst case, and an advantage over it
alone would overstate the contribution. The comparison must therefore also include a conventional
retrieval-augmented baseline, which generates every answer but uses no model for routing or
decomposition.

The emergency path is the most serious limitation of the current design. Keyword matching can miss
indirect phrasing, speech-recognition errors, and code-mixed expressions, and it can raise false
alarms, so it cannot serve as the principal safeguard. A production system would need, at a minimum,
a permanent one-tap emergency control, emergency numbers and instructions stored on the device,
explicit confirmation of the location and nature of the emergency, direct telephone and SMS paths,
escalation to a human operator, and a safe default whenever classification confidence is low. For
official one-to-many alerts, cell broadcast is preferable to any application-level channel, since it
is designed to remain effective under network congestion~\citep{3gpp23041}.

Several properties on which the design relies have yet to be measured. The accuracy of
nearest-neighbor routing on English, Hindi, Marathi, and code-mixed input is unknown; encoder quality
degrades with dialect and speech-recognition noise, and the thresholds that separate answering,
clarifying, deflecting, and escalating must be tuned and reported on a labeled multilingual set. It is
likewise an open question how small an InfoBin bundle can be while still giving useful offline
behavior, and what hit rate it then achieves. The cost argument of \cref{sec:cost} further assumes a
query distribution concentrated enough for the response cache to warm quickly, an assumption that only
real traffic can test.

Content currency poses a related problem. The device bundle is built at release time
(\cref{sec:semantic-cache}), and until signed incremental updates and revocation of stale or corrected
guidance are implemented, the offline tier cannot be relied on for advisories that change quickly. On
the server, guidance can become stale during road closures, shifts in crowd conditions, or
emergencies. The design provides last-updated labels, tiered refresh intervals, and revocation lists,
but the operational practice of keeping safety-critical InfoBins current, and the audit standards
appropriate to a public-sector deployment, remain to be established in the field.

Finally, the privacy posture described in \cref{sec:requirements} (bucketed rather than exact
attributes, local-first storage, coarse and consented location, aggregate dashboards by default, and
audit logs) is at present a design commitment. Whether it holds in deployment, and how much
personalization is worth its privacy cost, are questions for the field study.

\section{Conclusion}
\label{sec:conclusion}

Public information services at mass gatherings must serve very large, multilingual populations over
unreliable networks, under tight cost constraints, with some questions bearing directly on physical
safety. We have described \KD, an assistant for the Nashik Simhastha Kumbh Mela that treats retrieval
as the default and generation as a gated exception. A single retrieval primitive, cosine similarity
over partitions of an embedding-indexed store, provides the knowledge base, intent classifier, response
cache, offline bundle, and per-agent memory, and a hand-written orchestrator built on it calls a model
only when an explicit condition is met.

Our cost analysis shows that the advantage of this design over an LLM-default one is governed by the
share of traffic that must escalate to generation, a quantity we have argued is small for this domain
but have not yet measured. Measuring it, together with multilingual routing accuracy, offline coverage,
and the behavior of the emergency path, is the subject of our next phase of work. We expect the general
point to hold beyond this event: in bounded, safety-critical domains, generative models are most useful
when reserved for the queries that retrieval cannot answer.

\bibliographystyle{plainnat}
\bibliography{references}

\end{document}

%% file: figures/fig-runtime.tex
\def\lt{\sffamily\small\bfseries}\def\ld{\sffamily\footnotesize\color{kdink}}
\begin{tikzpicture}[
    font=\sffamily,
    layer/.style={
      draw=#1, fill=#1bg, line width=0.6pt, rounded corners=2.5pt,
      minimum width=10.6cm, minimum height=1.02cm,
      text width=10.2cm, align=left, inner sep=5pt, anchor=south west},
    lt/.style={font=\sffamily\small\bfseries},
    ld/.style={font=\sffamily\footnotesize, text=kdink},
  ]
  \def\step{1.24}
  \node[layer=kdgreen] (L0) at (0,0*\step)
    {{\lt\color{kdgreen!80!black}Curated content layer}\\[1pt]
     {\ld Official advisories, facilities, maps, schedules, emergency contacts}};
  \node[layer=kdgreen] (L1) at (0,1*\step)
    {{\lt\color{kdgreen!80!black}InfoBin layer}\\[1pt]
     {\ld Versioned, searchable knowledge units: metadata, semantics, safety rules}};
  \node[layer=kdgreen] (L2) at (0,2*\step)
    {{\lt\color{kdgreen!80!black}Local semantic cache (on device)}\\[1pt]
     {\ld Compact InfoBin bundle for fast lookup and offline behavior}};
  \node[layer=kdnavy] (L3) at (0,3*\step)
    {{\lt\color{kdnavy}Intent router}\\[1pt]
     {\ld Rules, lightweight NLP, cosine matching, confidence scoring, escalation}};
  \node[layer=kdnavy] (L4) at (0,4*\step)
    {{\lt\color{kdnavy}Bounded domain agents}\\[1pt]
     {\ld Food, travel, safety, accommodation, health, emergency, ritual, profile, crowd}};
  \node[layer=kdamber] (L5) at (0,5*\step)
    {{\lt\color{kdamber!85!black}Response composer}\\[1pt]
     {\ld Formats from approved content; small model or LLM only when required}};

  \draw[kdgrey, dashed, line width=0.6pt, rounded corners=2.5pt]
    (11.0,0) rectangle (12.55,5*\step+1.02);
  \node[rotate=90, align=center, font=\sffamily\small\bfseries, text=kdink]
    at (11.58,{(5*\step+1.02)/2}) {Observability layer};
  \node[rotate=90, align=center, font=\sffamily\scriptsize, text=kdgrey]
    at (12.07,{(5*\step+1.02)/2}) {latency $\cdot$ cost $\cdot$ cache hits $\cdot$ model use $\cdot$ sources $\cdot$ fallbacks};

  \draw[-{Stealth[length=6pt]}, kdgrey, line width=0.6pt] (-0.45,-0.15) -- (-0.45,5*\step+1.2);
  \node[rotate=90, font=\sffamily\scriptsize, text=kdgrey, anchor=south]
    at (-0.5,{(5*\step+1.02)/2}) {escalate only if needed};

  \begin{scope}[yshift=-0.55cm, font=\sffamily\scriptsize]
    \draw[kdgreen, fill=kdgreenbg, rounded corners=1pt] (0,0) rectangle +(0.28,-0.28);
    \node[anchor=west] at (0.36,-0.14) {zero generative LLM tokens};
    \draw[kdnavy, fill=kdnavybg, rounded corners=1pt] (3.9,0) rectangle +(0.28,-0.28);
    \node[anchor=west] at (4.26,-0.14) {deterministic; no LLM by default};
    \draw[kdamber, fill=kdamberbg, rounded corners=1pt] (8.2,0) rectangle +(0.28,-0.28);
    \node[anchor=west] at (8.56,-0.14) {may invoke a model};
  \end{scope}
\end{tikzpicture}

%% file: figures/fig-cache.tex
\begin{tikzpicture}[
    font=\sffamily,
    conv/.style={draw=kdred, fill=kdredbg, line width=0.6pt, rounded corners=2.5pt,
      minimum width=4.9cm, minimum height=0.78cm, font=\sffamily\footnotesize, text=kdred!80!black},
    chip/.style={draw=kdgold, fill=white, line width=0.5pt, rounded corners=2pt,
      minimum height=0.62cm, font=\sffamily\scriptsize, text=kdgold!70!black},
  ]
  \node[anchor=west, font=\sffamily\small\bfseries, text=kdred!85!black] at (0,0.75)
    {Conventional design: five separate systems};
  \foreach \i/\t in {0/Knowledge base (structured records),
                     1/Intent classifier (ML model or LLM),
                     2/Response cache (key--value),
                     3/Offline data bundle (static files),
                     4/Per-agent memory and state} {
    \node[conv, anchor=west] (c\i) at (0,-\i*1.02) {\t};
  }

  \draw[kdgold, fill=kdgoldbg, line width=0.8pt, rounded corners=4pt]
    (7.25,0.35) rectangle (14.05,-3.45);
  \node[font=\sffamily\small\bfseries, text=kdgold!65!black] at (10.65,-0.05) {Semantic cache};
  \node[font=\sffamily\scriptsize, text=kdgold!70!black] at (10.65,-0.5)
    {embedding-indexed records + metadata};
  \node[chip, minimum width=3.0cm] at (9.0,-1.2) {record embeddings};
  \node[chip, minimum width=3.0cm] at (12.3,-1.2) {intent exemplars};
  \node[chip, minimum width=3.0cm] at (9.0,-2.0) {cached Q\,$\to$\,A pairs};
  \node[chip, minimum width=3.0cm] at (12.3,-2.0) {POIs and facilities};
  \node[chip, minimum width=6.3cm] at (10.65,-2.8) {all queried by cosine similarity};

  \foreach \i/\y in {0/-0.45,1/-0.95,2/-1.45,3/-1.95,4/-2.45} {
    \draw[kdgrey, line width=0.5pt] (c\i.east) -- (7.25,\y);
  }
  \node[font=\sffamily\scriptsize\itshape, text=kdgrey, rotate=-6] at (6.05,0.12) {collapse into};

  \node[draw=kdink, fill=kdnavybg, line width=0.7pt, rounded corners=3pt, text=kdink,
        align=center, inner sep=6pt, minimum width=6.8cm] at (10.65,-4.2)
    {{\ttfamily\small cosine\_scores(query\_vec, partition)}\\[1pt]
     {\sffamily\scriptsize one operation; the partition determines the function}};
\end{tikzpicture}

%% file: figures/fig-ladder.tex
\def\ttl{\sffamily\small\bfseries}\def\dsc{\sffamily\scriptsize}
\begin{tikzpicture}[
    font=\sffamily,
    >={Stealth[length=5pt]},
    pill/.style={draw=kdink, fill=white, text=kdink, line width=0.7pt, rounded corners=9pt, align=center,
      inner xsep=12pt, inner ysep=5pt, font=\sffamily\small\bfseries},
    g/.style={draw=kdgreen, fill=kdgreenbg, line width=0.7pt, rounded corners=3pt,
      minimum width=6.3cm, minimum height=0.95cm, align=center, inner sep=4pt},
    a/.style={draw=kdamber, fill=kdamberbg, line width=0.7pt, rounded corners=3pt,
      minimum width=6.3cm, minimum height=0.95cm, align=center, inner sep=4pt},
    ttl/.style={font=\sffamily\small\bfseries},
    dsc/.style={font=\sffamily\scriptsize},
    flow/.style={->, line width=0.6pt, kdink},
    gflow/.style={->, line width=0.7pt, kdgreen},
    aflow/.style={->, line width=0.7pt, kdamber},
    lbl/.style={font=\sffamily\scriptsize, text=kdgrey},
  ]
  \node[pill] (q) at (0,0) {User query};
  \node[g] (s1) at (0,-1.45)
    {{\ttl\color{kdgreen!80!black}1 $\cdot$ Q\&A cache lookup (cosine)}\\
     {\dsc\color{kdgreen!80!black}paraphrase-aware, multilingual}};
  \node[g] (s2) at (0,-2.95)
    {{\ttl\color{kdgreen!80!black}2 $\cdot$ Intent routing (cosine $k$-NN)}\\
     {\dsc\color{kdgreen!80!black}$\approx$520 exemplars $\cdot$ 26 intents $\cdot$ emergency fast path}};
  \node[g] (s3) at (0,-4.45)
    {{\ttl\color{kdgreen!80!black}3 $\cdot$ Domain retrieval (hybrid)}\\
     {\dsc\color{kdgreen!80!black}FTS5 + cosine + rank fusion + rerank}};
  \node[diamond, aspect=2.6, draw=kdnavy, fill=kdnavybg, line width=0.7pt,
        align=center, inner sep=1pt] (d) at (0,-6.25)
    {{\ttl\color{kdnavy}Confidence}\\{\dsc\color{kdink}high or medium?}};
  \node[g, minimum width=4.2cm] (f) at (5.6,-6.25)
    {{\ttfamily\small\bfseries\color{kdgreen!80!black}format\_from\_records}\\
     {\dsc\color{kdgreen!80!black}grounded, cited $\cdot$ 0 LLM calls}};
  \node[a] (gate) at (0,-8.2)
    {{\ttl\color{kdamber!85!black}Escalation gate}\\
     {\dsc\color{kdamber!85!black}complex multi-part synthesis $\cdot$ low-confidence retrieval}\\[-1pt]
     {\dsc\color{kdamber!85!black}$\cdot$ live or external data}};
  \node[a, fill=kdamber!22, minimum width=5.2cm] (llm) at (0,-9.75)
    {{\ttl\color{kdamber!85!black}LLM (bounded fallback)}\\
     {\dsc\color{kdamber!85!black}constrained to retrieved records}};
  \node[a, minimum width=6.3cm] (gv) at (0,-11.15)
    {{\dsc\small\color{kdamber!85!black}Grounding validation: drop unsupported citations}};
  \node[pill, font=\sffamily\small] (ans) at (5.6,-9.75)
    {\textbf{Grounded answer}\\[-1pt]{\scriptsize\mdseries with source + status}};

  \draw[flow] (q) -- (s1);
  \draw[flow] (s1) -- node[lbl, right] {miss} (s2);
  \draw[flow] (s2) -- (s3);
  \draw[flow] (s3) -- (d);
  \draw[flow] (d) -- node[lbl, above] {yes} (f);
  \draw[flow] (d) -- node[lbl, right] {no / low} (gate);
  \draw[flow] (gate) -- (llm);
  \draw[flow] (llm) -- (gv);
  \draw[gflow] (f) -- (ans);
  \draw[aflow] (gv.east) -| (ans.south);
  \draw[gflow] (s1.east) -- node[lbl, above, text=kdgreen] {hit $\cdot$ 0 LLM calls}
    (8.15,-1.45) |- (ans.east);
  \node[lbl, rotate=-90, text=kdgreen, anchor=south] at (8.2,-5.6)
    {cache and cheap paths return here};
\end{tikzpicture}

%% file: figures/fig-tiers.tex
\begin{tikzpicture}[
    font=\sffamily,
    tier/.style={font=\sffamily\scriptsize\bfseries, text=kdgrey, anchor=west},
    orch/.style={draw=kdnavy, fill=kdnavybg, line width=0.8pt, rounded corners=3pt,
      align=center, inner sep=6pt, minimum width=6.6cm},
    agent/.style={draw=kdgreen, fill=kdgreenbg, line width=0.7pt, rounded corners=3pt,
      minimum width=2.75cm, minimum height=0.82cm, font=\sffamily\footnotesize\bfseries,
      text=kdgreen!75!black},
    subag/.style={draw=kdnavy!70, fill=kdnavybg!60, line width=0.5pt, rounded corners=2pt,
      minimum width=2.2cm, minimum height=0.8cm, align=center, font=\sffamily\scriptsize,
      text=kdink},
    part/.style={draw=kdgold, fill=white, line width=0.5pt, rounded corners=2pt,
      minimum width=2.45cm, minimum height=0.6cm, font=\sffamily\scriptsize,
      text=kdgold!70!black},
  ]
  \node[tier] at (0,0) {TIER 1};
  \node[tier] at (0,-2.0) {TIER 2};
  \node[tier] at (0,-3.75) {TIER 3};

  \node[orch] (O) at (8.6,0)
    {{\small\bfseries\color{kdnavy}Orchestrator}\\[1pt]
     {\scriptsize\color{kdink}decompose into sub-needs $\cdot$ dependency-wave scheduling}};

  \foreach \i/\x/\n in {1/3.4/Health, 2/6.85/Travel, 3/10.3/Commerce, 4/13.75/Crowd} {
    \node[agent] (A\i) at (\x,-2.0) {\n\ agent};
    \node[subag] (S\i) at (\x,-3.75) {InfoCard\\sub-agent};
    \draw[kdnavy, line width=0.6pt] (O.south) -- (A\i.north);
    \draw[kdnavy, line width=0.6pt] (A\i.south) -- (S\i.north);
  }

  \draw[kdgold, fill=kdgoldbg, line width=0.8pt, rounded corners=4pt]
    (1.75,-4.85) rectangle (15.45,-6.75);
  \node[font=\sffamily\small\bfseries, text=kdgold!65!black] at (8.6,-5.25)
    {Shared semantic cache (embedding-indexed, partitioned by domain)};
  \foreach \j/\x/\p in {1/3.15/health partition, 2/5.95/travel partition,
                        3/8.75/commerce partition, 4/11.55/crowd partition,
                        5/14.15/more partitions\,\dots} {
    \node[part] (P\j) at (\x,-6.15) {\p};
  }
  \foreach \i in {1,2,3,4} {
    \draw[kdgold, dashed, line width=0.6pt] (S\i.south) -- (P\i.north);
  }
\end{tikzpicture}

%% file: figures/fig-bands.tex
\begin{tikzpicture}[
    font=\sffamily,
    btitle/.style={font=\sffamily\normalsize\bfseries},
    bsub/.style={font=\sffamily\footnotesize, align=center, text width=4.2cm},
    bmech/.style={font=\sffamily\footnotesize, align=center, text width=4.2cm, text=kdink},
    bpill/.style={font=\sffamily\footnotesize\bfseries},
    bex/.style={font=\sffamily\scriptsize\itshape},
    bq/.style={font=\sffamily\footnotesize, align=center, text width=4.3cm, text=kdink},
  ]
  \draw[kdgrey, line width=0.5pt, -{Stealth[length=5pt]}] (0,5.45) -- (15.6,5.45);
  \node[anchor=south west, font=\sffamily\scriptsize, text=kdgrey] at (0,5.5)
    {lower cost, deterministic};
  \node[anchor=south east, font=\sffamily\scriptsize, text=kdgrey] at (15.6,5.5)
    {higher cost, generative};

  \foreach \x/\c/\cb/\B/\S/\M/\P/\Q in {
      2.5/kdgreen/kdgreenbg/A/{Similarity alone}/{Q\,$\to$\,A cache +\\direct retrieval}/{0 LLM calls}/%
        {``Where is Ramkund ghat?''\\``What are the snan dates?''},
      7.8/kdgold/kdgoldbg/B/{Similarity + coded\\orchestration}/{multi-record retrieval\\+ structured format}/{0 LLM calls}/%
        {``List hospitals near Panchavati\\with emergency care''},
      13.1/kdamber/kdamberbg/C/{LLM-backed synthesis\\and dialogue}/{multi-step planning,\\open-ended generation}/{1 or more LLM calls}/%
        {``I'm 72, diabetic, arriving\\Friday; plan everything for me''}} {
    \draw[\c, fill=\cb, line width=0.8pt, rounded corners=4pt]
      (\x-2.5,-0.25) rectangle (\x+2.5,4.75);
    \node[btitle, text=\c!80!black] at (\x,4.3) {Band \B};
    \node[bsub, text=\c!80!black] at (\x,3.55) {\S};
    \draw[\c!70, line width=0.5pt] (\x-1.9,2.95) -- (\x+1.9,2.95);
    \node[bmech] at (\x,2.4) {\M};
    \node[bpill, text=\c!85!black] at (\x,1.45) {\P};
    \node[bex, text=\c!80!black] at (\x,0.8) {Example};
    \node[bq] at (\x,0.25) {\Q};
  }

  \node[font=\sffamily\scriptsize, text=kdink, align=center] at (7.8,-0.75)
    {Outside the reach of Bands A and B: content absent from the corpus, multi-step reasoning,
     novel synthesis, and transactions that require tools};
\end{tikzpicture}

%% file: references.bib
@article{sparckjones1972statistical,
  author  = {Sp{\"a}rck Jones, Karen},
  title   = {A Statistical Interpretation of Term Specificity and Its Application in Retrieval},
  journal = {Journal of Documentation},
  volume  = {28},
  number  = {1},
  pages   = {11--21},
  year    = {1972},
  doi     = {10.1108/eb026526}
}

@article{salton1975vector,
  author  = {Salton, Gerard and Wong, Anita and Yang, Chung-Shu},
  title   = {A Vector Space Model for Automatic Indexing},
  journal = {Communications of the ACM},
  volume  = {18},
  number  = {11},
  pages   = {613--620},
  year    = {1975},
  doi     = {10.1145/361219.361220}
}

@book{manning2008introduction,
  author    = {Manning, Christopher D. and Raghavan, Prabhakar and Sch{\"u}tze, Hinrich},
  title     = {Introduction to Information Retrieval},
  publisher = {Cambridge University Press},
  address   = {Cambridge, UK},
  year      = {2008},
  doi       = {10.1017/CBO9780511809071}
}

@article{deerwester1990indexing,
  author  = {Deerwester, Scott and Dumais, Susan T. and Furnas, George W. and Landauer, Thomas K. and Harshman, Richard},
  title   = {Indexing by Latent Semantic Analysis},
  journal = {Journal of the American Society for Information Science},
  volume  = {41},
  number  = {6},
  pages   = {391--407},
  year    = {1990},
  doi     = {10.1002/(SICI)1097-4571(199009)41:6<391::AID-ASI1>3.0.CO;2-9}
}

@misc{mikolov2013efficient,
  author        = {Mikolov, Tomas and Chen, Kai and Corrado, Greg and Dean, Jeffrey},
  title         = {Efficient Estimation of Word Representations in Vector Space},
  year          = {2013},
  eprint        = {1301.3781},
  archivePrefix = {arXiv},
  primaryClass  = {cs.CL},
  howpublished  = {arXiv preprint arXiv:1301.3781}
}

@inproceedings{reimers2019sentence,
  author    = {Reimers, Nils and Gurevych, Iryna},
  title     = {{Sentence-BERT}: Sentence Embeddings using {S}iamese {BERT}-Networks},
  booktitle = {Proceedings of the 2019 Conference on Empirical Methods in Natural Language Processing and the 9th International Joint Conference on Natural Language Processing (EMNLP-IJCNLP)},
  pages     = {3982--3992},
  year      = {2019},
  doi       = {10.18653/v1/D19-1410}
}

@misc{cheng2026enhancing,
  author        = {Cheng, Zhiyuan and Lai, Longying and Liu, Yue and Cheng, Kai and Qi, Xiaoxi},
  title         = {Enhancing Financial Report Question-Answering: A Retrieval-Augmented Generation System with Reranking Analysis},
  year          = {2026},
  eprint        = {2603.16877},
  archivePrefix = {arXiv},
  howpublished  = {arXiv preprint arXiv:2603.16877}
}

@inproceedings{lewis2020retrieval,
  author    = {Lewis, Patrick and Perez, Ethan and Piktus, Aleksandra and Petroni, Fabio and Karpukhin, Vladimir and Goyal, Naman and K{\"u}ttler, Heinrich and Lewis, Mike and Yih, Wen-tau and Rockt{\"a}schel, Tim and Riedel, Sebastian and Kiela, Douwe},
  title     = {Retrieval-Augmented Generation for Knowledge-Intensive {NLP} Tasks},
  booktitle = {Advances in Neural Information Processing Systems},
  volume    = {33},
  pages     = {9459--9474},
  year      = {2020}
}

@misc{yao2022react,
  author        = {Yao, Shunyu and Zhao, Jeffrey and Yu, Dian and Du, Nan and Shafran, Izhak and Narasimhan, Karthik and Cao, Yuan},
  title         = {{ReAct}: Synergizing Reasoning and Acting in Language Models},
  year          = {2022},
  eprint        = {2210.03629},
  archivePrefix = {arXiv},
  primaryClass  = {cs.CL},
  howpublished  = {arXiv preprint arXiv:2210.03629}
}

@misc{hinton2015distilling,
  author        = {Hinton, Geoffrey and Vinyals, Oriol and Dean, Jeff},
  title         = {Distilling the Knowledge in a Neural Network},
  year          = {2015},
  eprint        = {1503.02531},
  archivePrefix = {arXiv},
  primaryClass  = {stat.ML},
  howpublished  = {arXiv preprint arXiv:1503.02531}
}

@misc{mcmahan2016communication,
  author        = {McMahan, H. Brendan and Moore, Eider and Ramage, Daniel and Hampson, Seth and Ag{\"u}era y Arcas, Blaise},
  title         = {Communication-Efficient Learning of Deep Networks from Decentralized Data},
  year          = {2016},
  eprint        = {1602.05629},
  archivePrefix = {arXiv},
  primaryClass  = {cs.LG},
  howpublished  = {arXiv preprint arXiv:1602.05629}
}

@misc{han2015deep,
  author        = {Han, Song and Mao, Huizi and Dally, William J.},
  title         = {Deep Compression: Compressing Deep Neural Networks with Pruning, Trained Quantization and {H}uffman Coding},
  year          = {2015},
  eprint        = {1510.00149},
  archivePrefix = {arXiv},
  primaryClass  = {cs.CV},
  howpublished  = {arXiv preprint arXiv:1510.00149}
}

@inproceedings{brown2020language,
  author    = {Brown, Tom and Mann, Benjamin and Ryder, Nick and Subbiah, Melanie and Kaplan, Jared D. and Dhariwal, Prafulla and Neelakantan, Arvind and Shyam, Pranav and Sastry, Girish and Askell, Amanda and others},
  title     = {Language Models are Few-Shot Learners},
  booktitle = {Advances in Neural Information Processing Systems},
  volume    = {33},
  pages     = {1877--1901},
  year      = {2020}
}

@inproceedings{guu2020retrieval,
  author    = {Guu, Kelvin and Lee, Kenton and Tung, Zora and Pasupat, Panupong and Chang, Mingwei},
  title     = {Retrieval Augmented Language Model Pre-Training},
  booktitle = {Proceedings of the 37th International Conference on Machine Learning},
  series    = {Proceedings of Machine Learning Research},
  volume    = {119},
  pages     = {3929--3938},
  publisher = {PMLR},
  year      = {2020}
}

@inproceedings{karpukhin2020dense,
  author    = {Karpukhin, Vladimir and Oguz, Barlas and Min, Sewon and Lewis, Patrick and Wu, Ledell and Edunov, Sergey and Chen, Danqi and Yih, Wen-tau},
  title     = {Dense Passage Retrieval for Open-Domain Question Answering},
  booktitle = {Proceedings of the 2020 Conference on Empirical Methods in Natural Language Processing (EMNLP)},
  pages     = {6769--6781},
  year      = {2020},
  doi       = {10.18653/v1/2020.emnlp-main.550}
}

@inproceedings{shuster2021retrieval,
  author    = {Shuster, Kurt and Poff, Spencer and Chen, Moya and Kiela, Douwe and Weston, Jason},
  title     = {Retrieval Augmentation Reduces Hallucination in Conversation},
  booktitle = {Findings of the Association for Computational Linguistics: EMNLP 2021},
  pages     = {3784--3803},
  year      = {2021},
  doi       = {10.18653/v1/2021.findings-emnlp.320}
}

@inproceedings{gao2021simcse,
  author    = {Gao, Tianyu and Yao, Xingcheng and Chen, Danqi},
  title     = {{SimCSE}: Simple Contrastive Learning of Sentence Embeddings},
  booktitle = {Proceedings of the 2021 Conference on Empirical Methods in Natural Language Processing},
  pages     = {6894--6910},
  year      = {2021},
  doi       = {10.18653/v1/2021.emnlp-main.552}
}

@inproceedings{cormack2009reciprocal,
  author    = {Cormack, Gordon V. and Clarke, Charles L. A. and B{\"u}ttcher, Stefan},
  title     = {Reciprocal Rank Fusion Outperforms {C}ondorcet and Individual Rank Learning Methods},
  booktitle = {Proceedings of the 32nd International ACM SIGIR Conference on Research and Development in Information Retrieval},
  pages     = {758--759},
  year      = {2009},
  doi       = {10.1145/1571941.1572114}
}

@misc{nogueira2019passage,
  author        = {Nogueira, Rodrigo and Cho, Kyunghyun},
  title         = {Passage Re-ranking with {BERT}},
  year          = {2019},
  eprint        = {1901.04085},
  archivePrefix = {arXiv},
  primaryClass  = {cs.IR},
  howpublished  = {arXiv preprint arXiv:1901.04085}
}

@techreport{3gpp23041,
  author      = {{3rd Generation Partnership Project}},
  title       = {Technical Realization of Cell Broadcast Service ({CBS})},
  institution = {3GPP},
  type        = {Technical Specification},
  number      = {TS 23.041},
  year        = {2024},
  note        = {Release 18}
}
